\documentclass{article}
\usepackage{graphicx}
\usepackage{glas}

\begin{document}

\begin{titlepage}{GLAS-PPE/2009-08}{15$^{\underline{\rm{th}}}$ May 2009}

\title{Optimised access to user analysis data using the gLite DPM}

\author{Ê 
Sam Skipsey $^1$, Greig Cowan $^2$ , Mike Kenyon $^1$,\\ 
 Stuart Purdie $^1$, Graeme Stewart $^1$\\
\\
$^1$ University of Glasgow, Glasgow, G12 8QQ, Scotland\\
$^2$ Department of Physics, University of Edinburgh, Edinburgh, EH9 3JZ}


\begin{abstract}
The ScotGrid distributed Tier-2 now provides more that 4MSI2K and 500TB for LHC computing, which is spread across three sites at Durham, Edinburgh and Glasgow. Tier-2 sites have a dual role to play in the computing models of the LHC VOs. Firstly, their CPU resources are used for the generation of Monte Carlo event data. Secondly, the end user analysis data is distributed across the grid to the site's storage system and held on disk ready for processing by physicists' analysis jobs. In this paper we show how we have designed the ScotGrid storage and data management resources in order to optimise access by physicists to LHC data. Within ScotGrid, all sites use the gLite DPM storage manager middleware. Using the EGEE grid to submit real ATLAS analysis code to process VO data stored on the ScotGrid sites, we present an analysis of the performance of the architecture at one site, and procedures that may be undertaken to improve such. The results will be presented from the point of view of the end user (in terms of number of events processed/second) and from the point of view of the site, which wishes to minimise load and the impact that analysis activity has on other users of the system.
\vspace{0.5cm}
\begin{center}
{\em 17$^{\underline{\rm{th}}}$ International Conference on Computing in High Energy and Nuclear Physics}\\
{\em Prague, Czech Republic}
\end{center}
\end{abstract}

\newpage
\end{titlepage}

\section{Introduction}
In general, WLCG VOs (ATLAS, CMS, LHCb and ALICE) have two main uses for supporting sites: \emph{Monte Carlo production}, in which simulation data is produced for use in event selection and analysis, and \emph{user analysis}, in which real detector data from the LHC is analysed\footnote{We note here that the LHCb computing model foresees analysis only at Tier-1 centres and reserves Tier-2 CPU capacity for simulation}. 
Production jobs have run on the Grid for several years, building up stocks of simulation data in preparation for the LHC turn on. As a result, sites and experiments have optimised their existing infrastructure and best practice for production work. Sites supporting ATLAS Production, for example, have well-understood requirements on data storage and transfer required for a given total computational power, and procure (within constraints from other VOs) to provide infrastructure matching those requirements.
User analysis is, by comparison, poorly understood, mainly because ``true'' user analysis is contingent on the generation of actual data at the LHC. It is widely accepted that better understanding of the behavior of such jobs is essential to providing best practice and infrastructure recommendations for sites, and for experiments to properly provide advice to their members. 

The ATLAS VO has begun test simulations of user analysis patterns against Tier-2 sites in order to probe their performance against this kind of workload. The ``HammerCloud''\cite{hammercloud} framework, which utlises the Ganga\cite{ganga} Python-based grid user interface to automate job submission and statistics calculation, is capable of submitting hundreds of analysis jobs (based on a single real analysis workflow) against a site or group of sites. 
UKI-SCOTGRID-GLASGOW, the Glasgow Tier-2 site, has used regular HammerCloud tests, consisting of approximately 300 jobs per run, to test the performance of their storage infrastructure against user analysis load, and to monitor the effect of improvements made.

Glasgow's storage infrastructure, before optimisation, consisted of 
\begin{itemize}
\item DPM\cite{dpm} head node (svr018.gla.scotgrid.ac.uk) running on a dual-core 2.5GHz AMD Opteron with 8Gb of RAM. The MySQL database backend for the DPM services was hosted on 2 Hitachi 10,000RPM Ultrastar harddisks in a RAID 1 configuration, in order to provide fast, low-latency access with some data security. 
\item 18 disk pool servers in a mix of configurations. The majority (newer) disks have 20 SATA disks with a PCI-X hardware controller, configured in a single RAID 6 array and partitioned across 5 volumes.
\item Gigabit ethernet links between all services.
 \end{itemize}
This configuration was more than sufficient to support ATLAS production work, never achieving high load on the head node nor the disk servers even with a full cluster (of 1920) jobs.

\section{HammerCloud results against unoptimised configuration}
HammerCloud test 38 (hence referred to as HC38) is a typical HammerCloud test run against Glasgow before any optimisations were performed. Figure \ref{fig1} shows the CPU load on the DPM head node approaching 100\%, clearly bottlenecking the performance of the rest of the infrastructure. By comparison, the disk pool nodes are barely loaded. The HammerCloud statistics (figure \ref{fig2}) show that we achieved an event-rate of almost 10Hz.

Initial testing showed that performance of the DPM storage was noticably lacking, barely achieving event-rates of 10Hz, with a concomitant limitation on the mean and maximum job efficiency. The load on the DPM head node itself implies that it is clearly the effective bottleneck in this case.

\begin{figure}
\begin{minipage}{18pc}
\includegraphics[width=18pc]{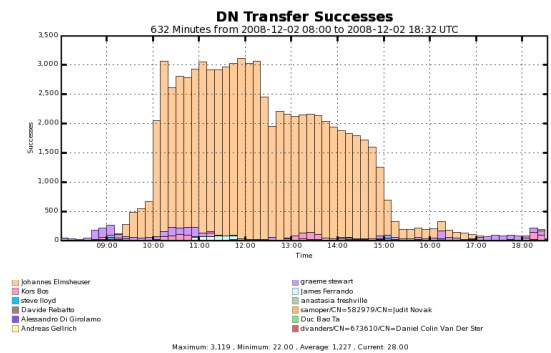}
a) Successful transfers (600 second intervals) by DN
\end{minipage}
\hspace{2pc}
\begin{minipage}{18pc}
\includegraphics[width=18pc]{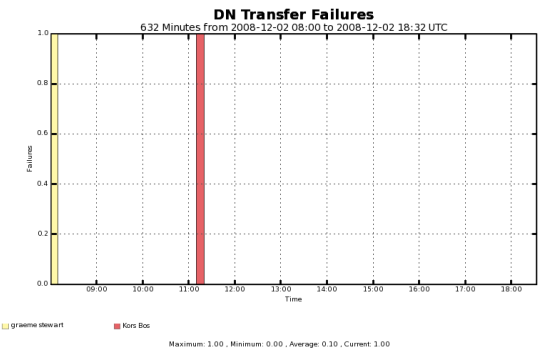}
b) Failed transfers (600 second intervals) by DN
\end{minipage}
\caption{\label{fig0} DPM statistics for HammerCloud test ``HC38''.}
\end{figure}
\begin{figure}
\begin{minipage}{38pc}
\centering
\begin{minipage}{18pc}
\includegraphics[width=18pc]{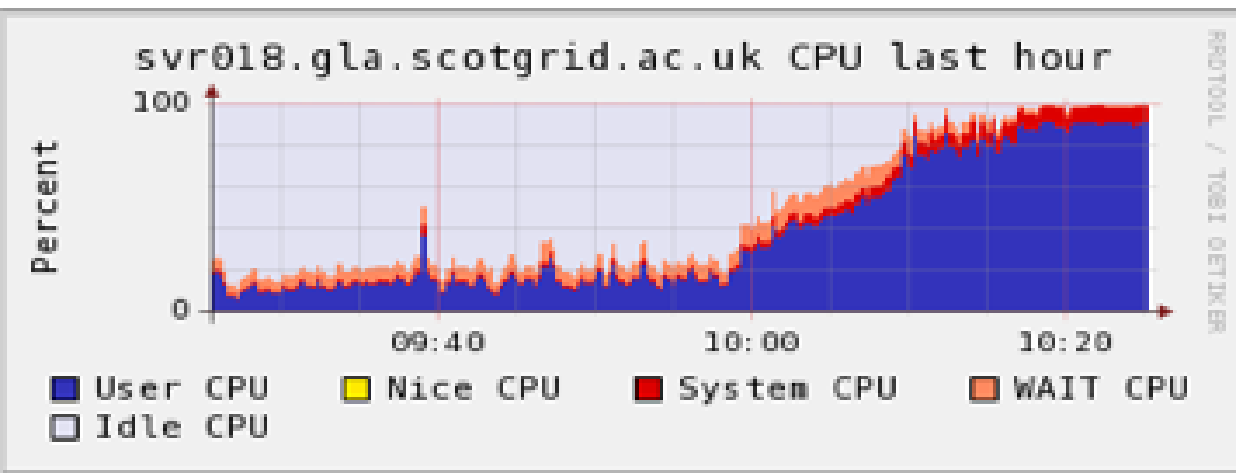}
a) CPU load on DPM Head node (percentage of full load)
\end{minipage}
\end{minipage}
\begin{minipage}{38pc}
\begin{minipage}{18pc}
\includegraphics[width=18pc]{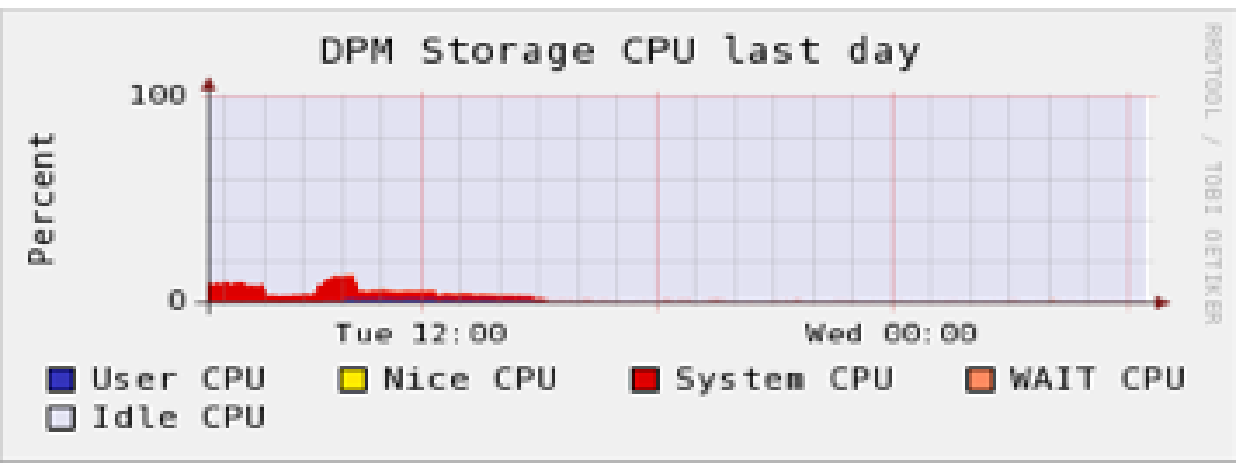}
b) CPU load on DPM pool (average, percentage of full load)
\end{minipage}
\hspace{2pc}
\begin{minipage}{18pc}
\includegraphics[width=18pc]{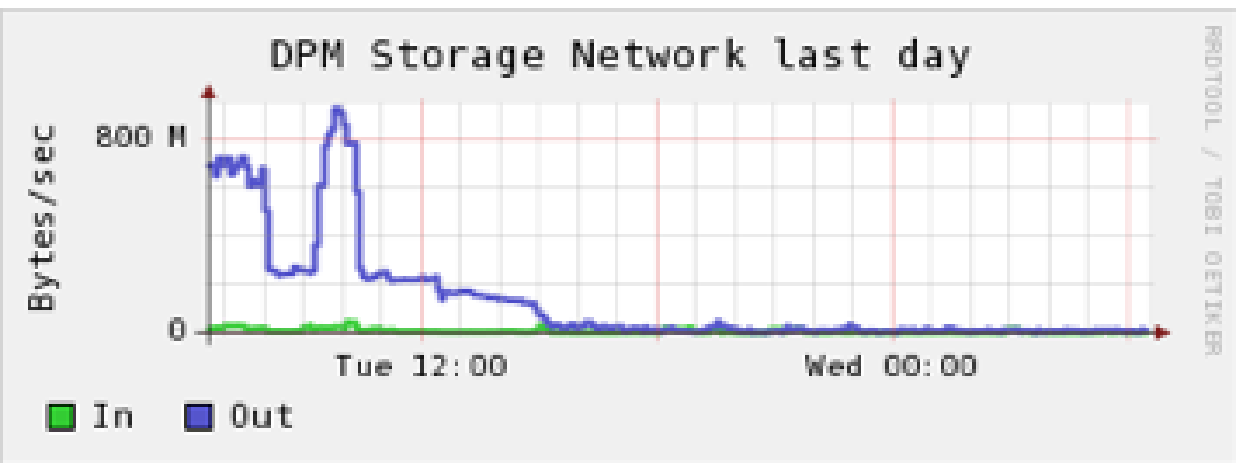}
c) Network load on DPM pool (total, bytes/sec)
\end{minipage}
\end{minipage}
\caption{\label{fig1} Load on services during HammerCloud test ``HC38''.}
\end{figure}
\begin{figure}
\begin{minipage}{10pc}
\includegraphics[width=10pc]{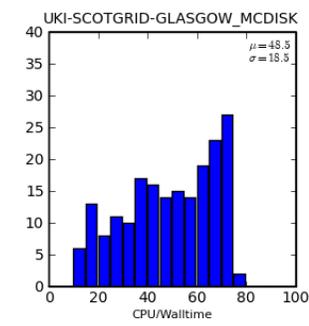}
a) CPU efficiency (cputime/walltime)
\end{minipage}
\hspace{2pc}
\begin{minipage}{10pc}
\includegraphics[width=10pc]{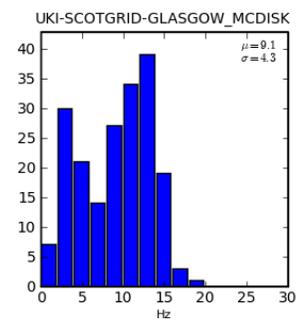}
b) Event rate (events per second per job)
\end{minipage}
\hspace{2pc}
\begin{minipage}{12pc}
\includegraphics[width=12pc]{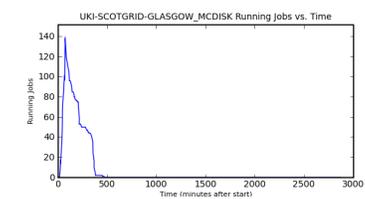}
c) Simultaneous running jobs
\end{minipage}
\caption{\label{fig2} Statistics generated by HammerCloud test ``HC38'' for the Glasgow site}
\end{figure}

\section{Optimisation Steps}
\subsection{Service separation}
As the bottleneck appeared to be the available computational power on the head node, we considered moving the DPM services to newer hardware. However, on investigation of the load pattern on the head node, we determined that all the CPU load was due to the ``srmv2.2'' and ``dpm'' processes, whilst the MySQL backend was engaging in large amounts of IO activity. Considering the limitations on available hardware, we decided to adopt a split head node configuration: the MySQL database would remain on the old hardware, to take advantage of the fast disks there, whilst the DPM services would be moved onto a repurposed worker node.
In the process of this move we renamed the machines to keep the DPM services associated with the same DNS name. Hence, ``svr015.gla.scotgrid.ac.uk'' is the new name for the original hardware, and the worker node ``node310'' was renamed ``svr018.gla.scotgrid.ac.uk''.
The partitioning process, including suitable reimaging of the worker node, and reconfiguration of the services, took less than 2 hours, with the significant advantage of Glasgow's cfengine-based cluster configuration management system.
The new configuration of the DPM head was:
\begin{itemize}
\item MySQL server on old head node dual-core Opteron, 8GB memory with 2 10,00RPM disk RAID 1 array
\item DPM services on dual quad-core 2.5GHz Intel Xeon with 16Gb memory, small 7200RPM "energy efficient" hard disk
\end{itemize}

\subsection{Post-separation tests}
\begin{figure}
\begin{minipage}{18pc}
\includegraphics[width=18pc]{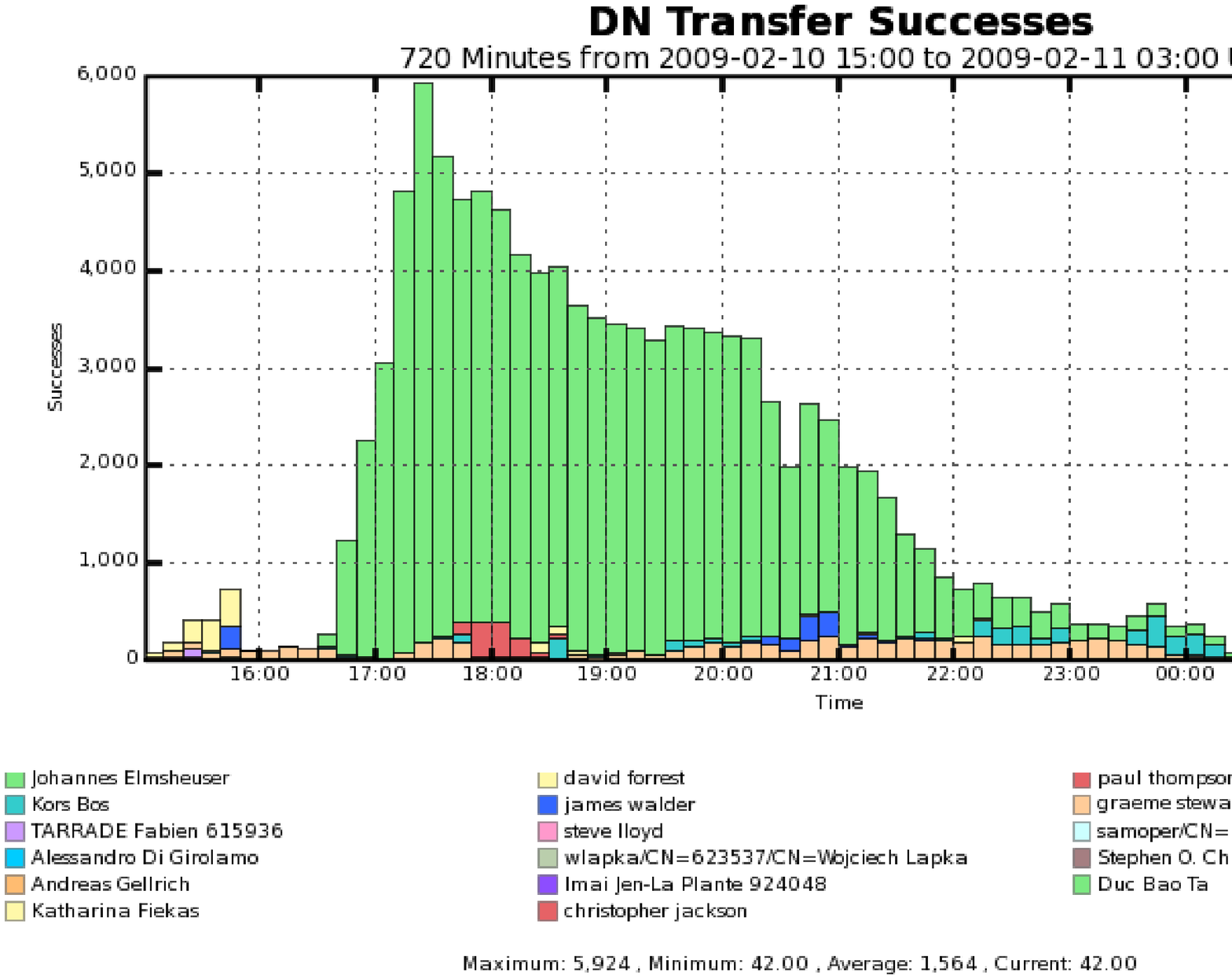}
a) Successful transfers (600 second intervals) by DN
\end{minipage}
\hspace{2pc}
\begin{minipage}{18pc}
\includegraphics[width=18pc]{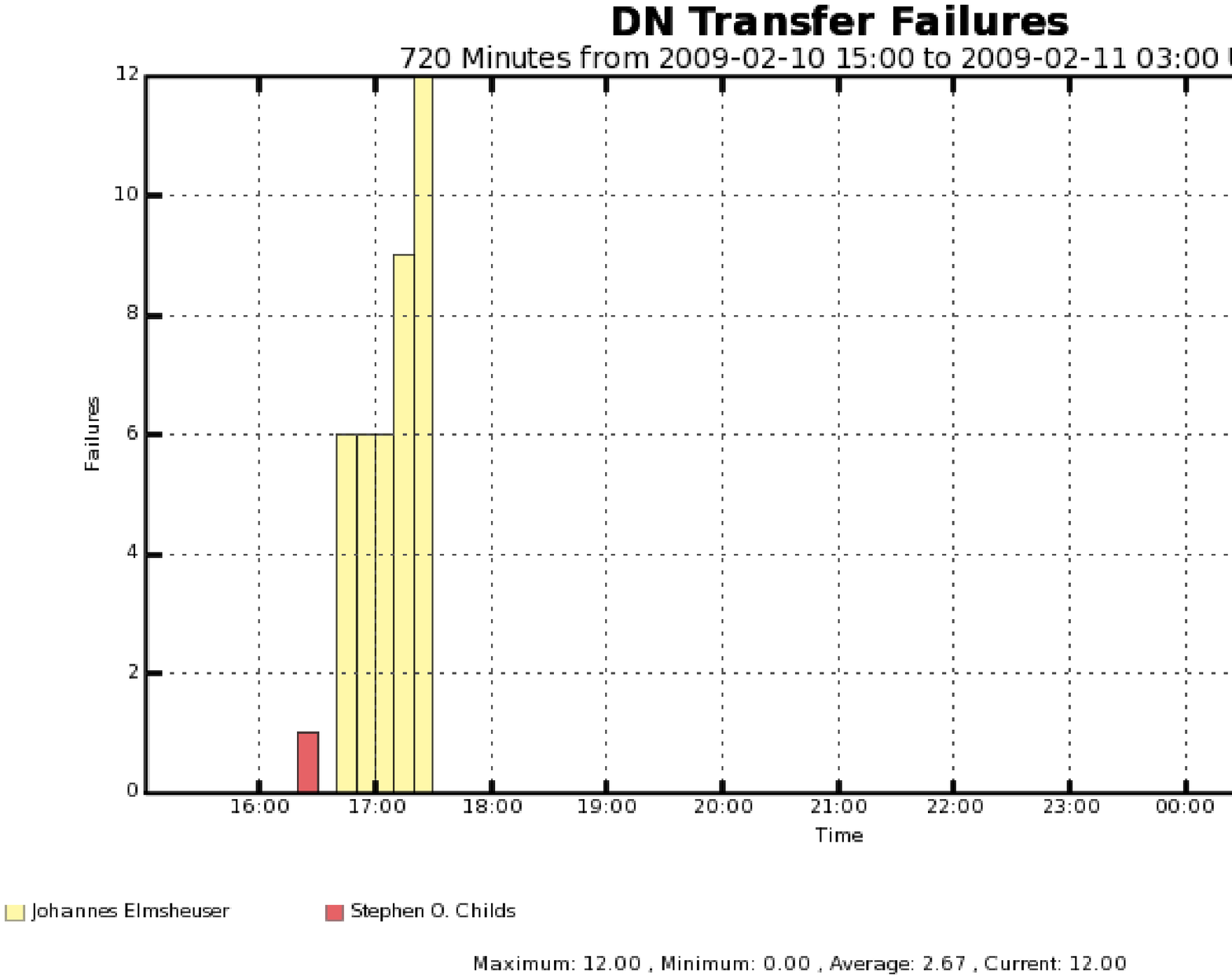}
b) Failed transfers (600 second intervals) by DN
\end{minipage}
\caption{\label{fig3a} DPM statistics for HammerCloud test ``HC135''.}
\end{figure}

\begin{figure}
\begin{minipage}{38pc}
\begin{minipage}{18pc}
\includegraphics[width=18pc]{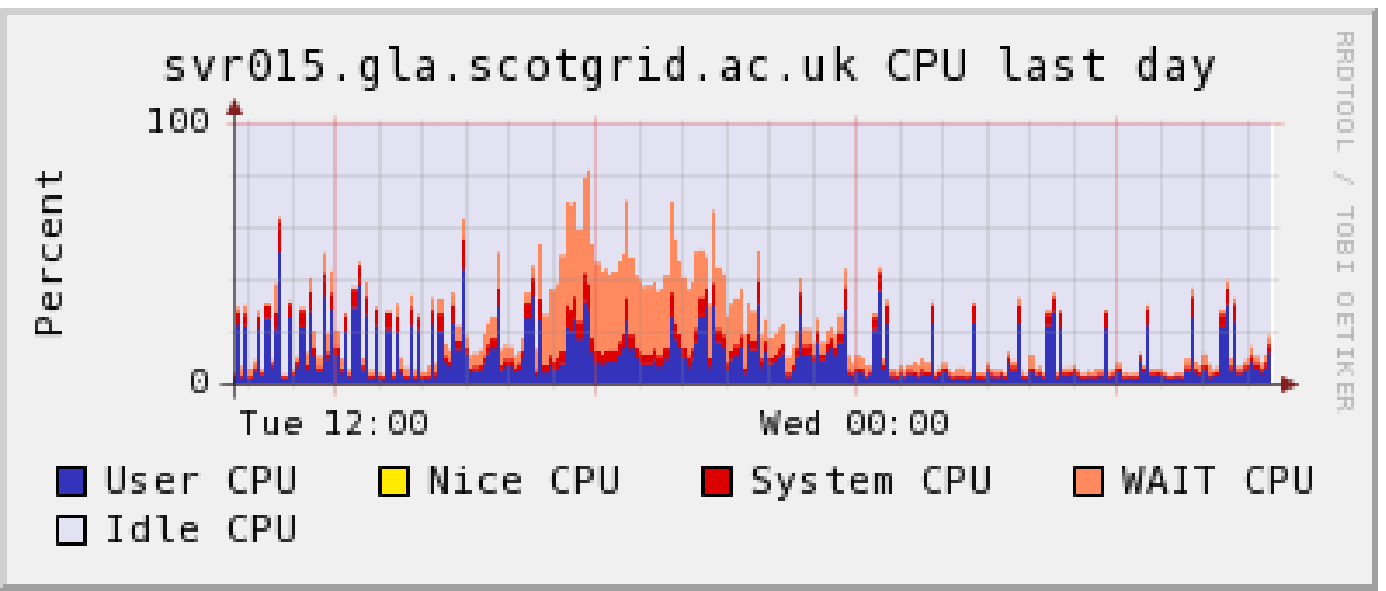}
a) CPU load on MySQL server (percentage of full load)
\end{minipage}
\hspace{2pc}
\begin{minipage}{18pc}
\includegraphics[width=18pc]{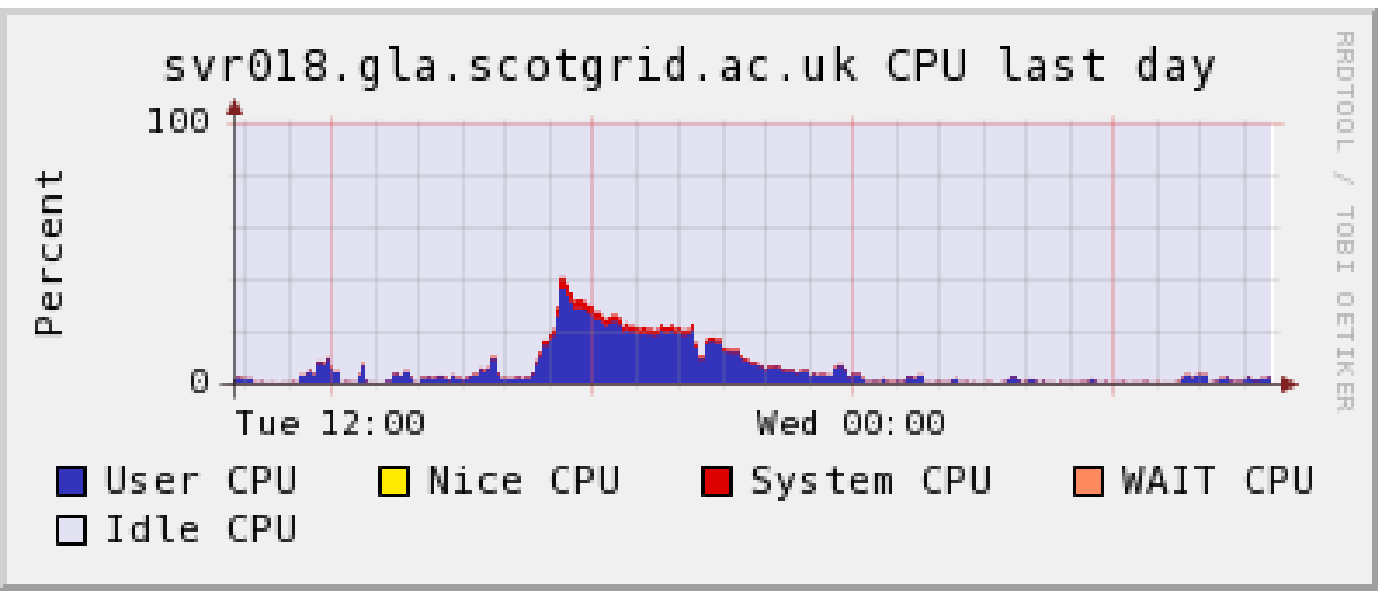}
b) CPU load on DPM services node (percentage of full load)
\end{minipage}
\end{minipage}
\begin{minipage}{38pc}
\begin{minipage}{18pc}
\includegraphics[width=18pc]{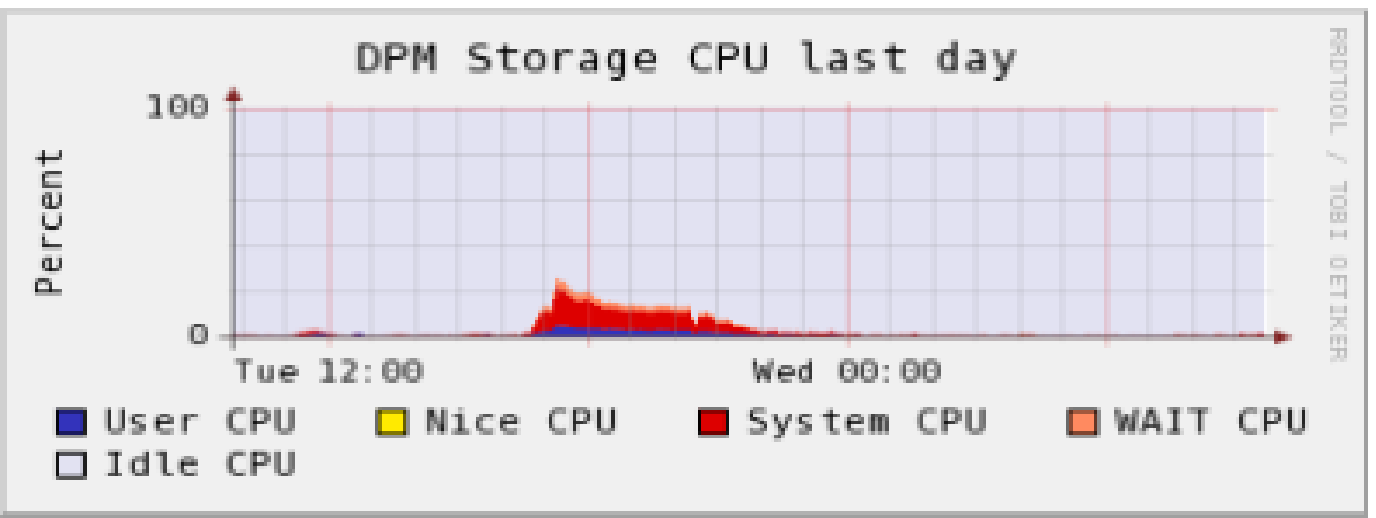}
c) CPU load on DPM pool (average, percentage of full load)
\end{minipage}
\hspace{2pc}
\begin{minipage}{18pc}
\includegraphics[width=18pc]{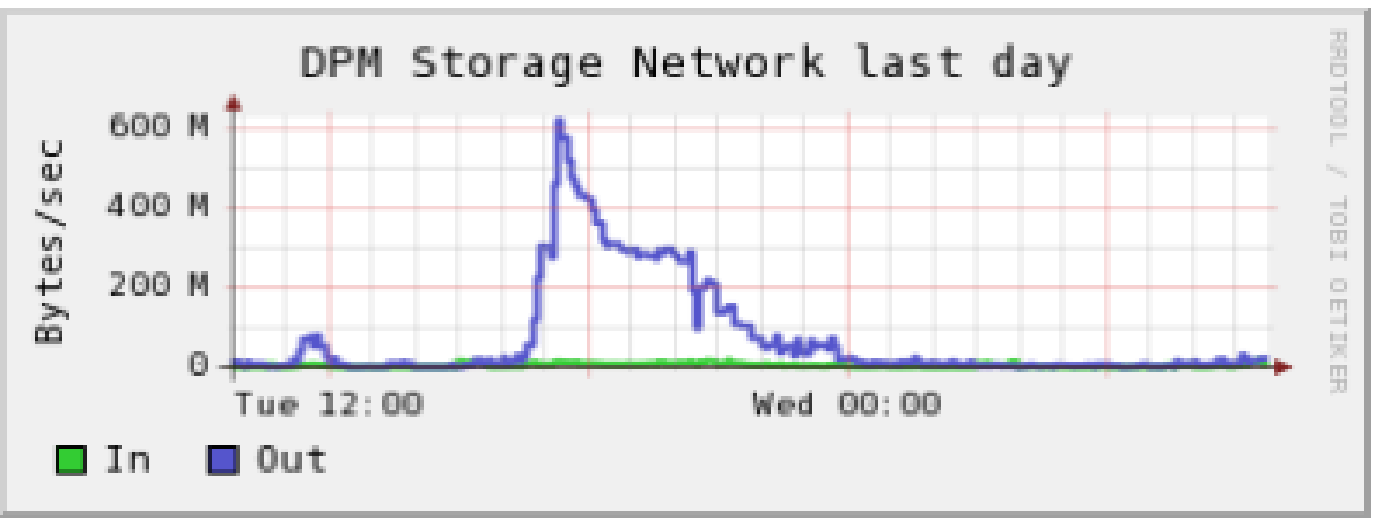}
d) Network load on DPM pool (total, bytes/sec)
\end{minipage}
\end{minipage}
\caption{\label{fig3} Load on services during HammerCloud test ``HC135''.}
\end{figure}

\begin{figure}
\begin{minipage}{10pc}
\includegraphics[width=10pc]{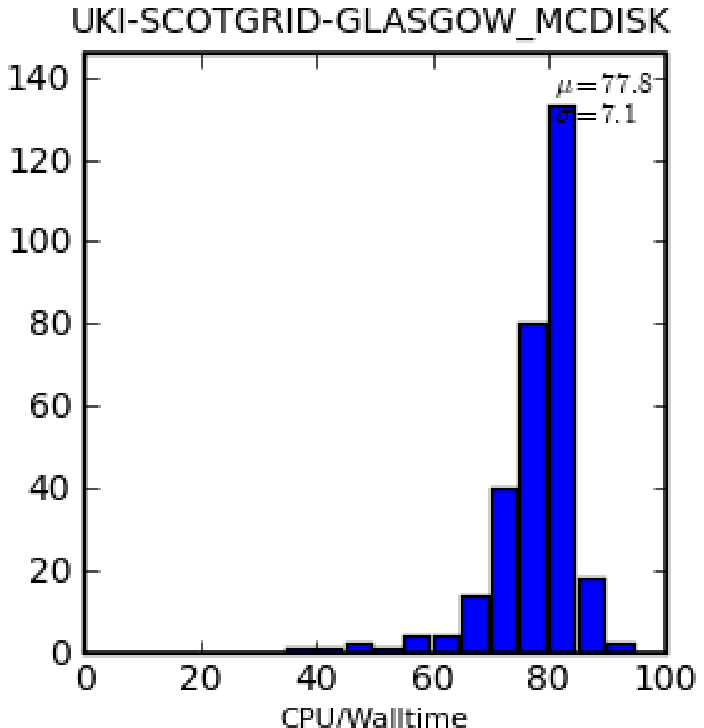}
a) CPU efficiency (cputime/walltime)
\end{minipage}
\hspace{2pc}
\begin{minipage}{10pc}
\includegraphics[width=10pc]{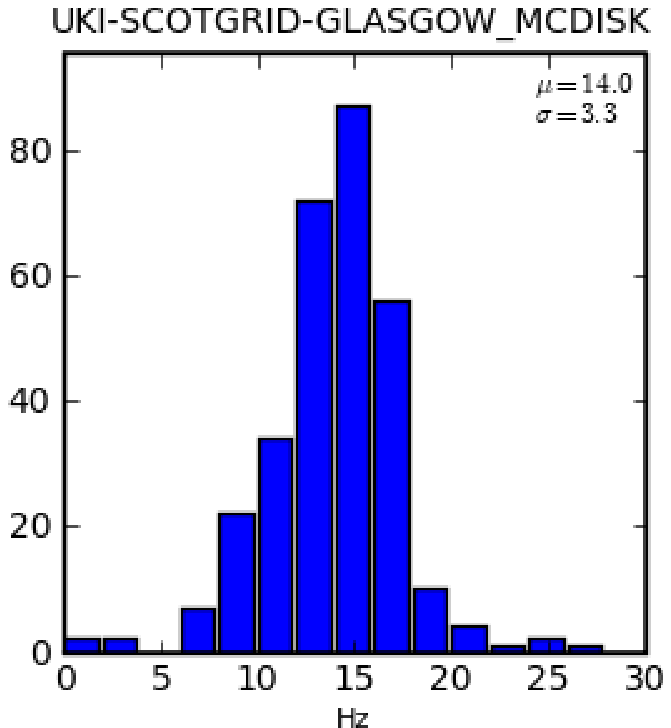}
b) Event rate (events per second per job)
\end{minipage}
\hspace{2pc}
\begin{minipage}{12pc}
\includegraphics[width=12pc]{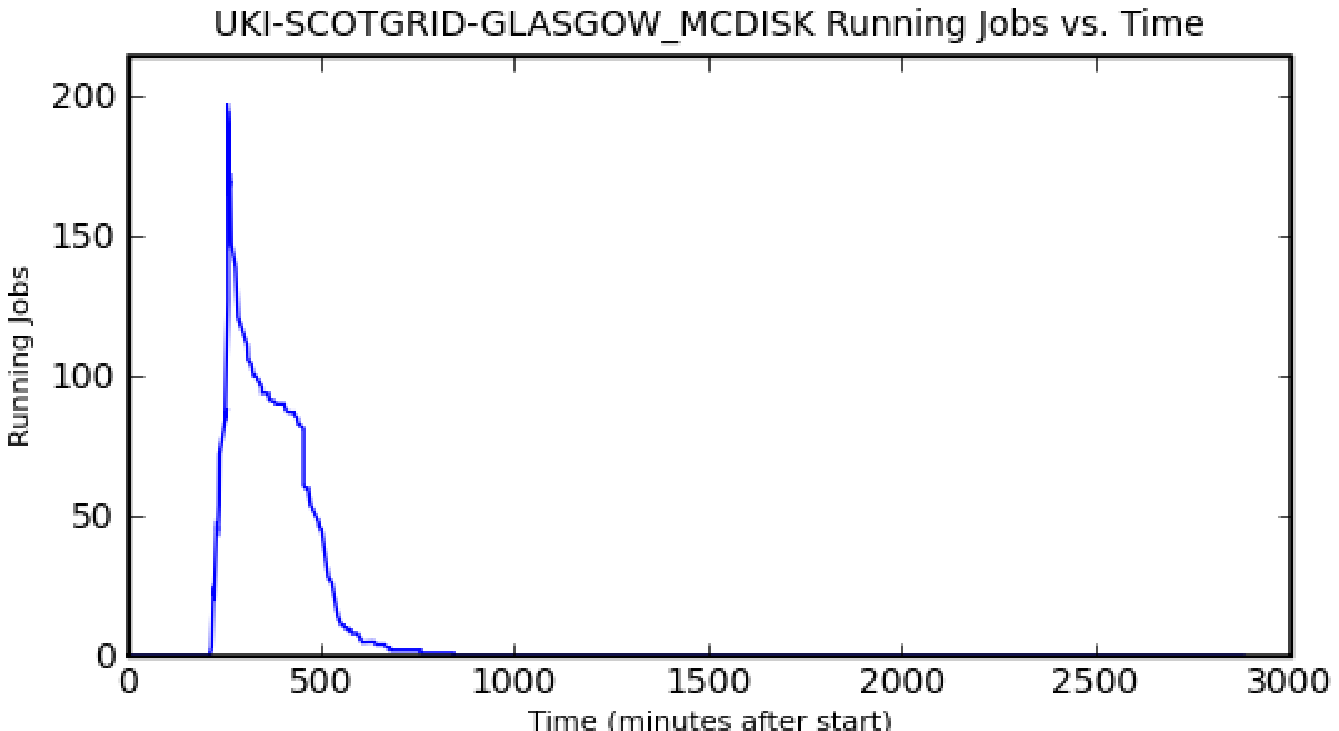}
c) Simultaneous running jobs
\end{minipage}
\caption{\label{fig4} Statistics generated by HammerCloud test ``HC135'' for the Glasgow site.}
\end{figure}

The next HammerCloud test, HC135, ran against the reconfigured hardware, producing the results in figures \ref{fig3a} to \ref{fig4}. As can be seen, the cpu load on the DPM head node is roughly comparable to the previous load, considering the quadrupling of the head node's compute power. As before, the disk pool servers are barely loaded; however, the MySQL server component now has considerable IOWait showing.
With this reconfiguration, our event-rate increased by 40\%, to around 14Hz, with a concomitant increase in our job efficiency (as most of the inefficiency of an analysis job is caused by waits for data to arrive). This is also reflected in the doubling of the successful transfers logged by DPM. However, as the pool servers were still relatively unloaded, it seemed reasonable to assume that further efficiency gains could be achieved with optimisations on the MySQL server. 

\subsection{MySQL optimisation}
\subsubsection{Building indices on common queries}

Analysis of the points of stress by enabling slow query logging in MySQL showed that the most common queries which increased load were on unindexed, or non-optimally indexed, quantities in tables mainly in the \texttt{dpm\_db} used by DPM to manage requests. The SRM\cite{srm} protocol requires that a storage system manage all requests in a resilient and stateful manner, thus resulting in each transfer request to the head node causing writes to an appropriate table (\texttt{dpm\_get\_filereq} for get requests, for example, which are the most common requests against a Tier-2 storage system in general), and additional writes on each completion. 
Whilst the tables used by DPM have generally good indexing, we identified a few cases which benefited from additional index creation.

We added the following index to the database:

\begin{verbatim}
create index pfn_lifetime on dpm_get_filereq (pfn(255), lifetime); 
\end{verbatim}

this actually modifies an existing index (only on pfn) to a composite index on pfn, lifetime. The associated queries are extremely common on DPMs servicing data requests, and so the small gain in efficiency from the compound index adds up over the large number of such queries produced by HammerCloud style user analysis.

We also discovered that monitoring tools deployed on our DPM produced spikes of high IO load when querying the request tables. The slow queries log again provided hints as to the guilty queries, allowing us to remove most of the load by adding the indices: 

\begin{verbatim}
create index status_idx on dpm_put_filereq(status);
create index stime_idx on dpm_req(stime); 
\end{verbatim}

Finally, and related to the above, we determined that the MonAMI\cite{monami} DPM plugin was performing fairly frequent queries against the \emph{cns\_db} (which stores the DPM namespace, and thus information about all the files stored) in order to obtain filesystem information. Whilst the \emph{cns\_db} is extremely well indexed, we determined that in this case, the addition of 
\begin{verbatim}
create index usage_by_group Cns_file_metadata(gid, filesize); 
\end{verbatim}
removed almost all the load produced by MonAMI's frequent global queries. Most of the ``noise'' visible in the load plot for the MySQL server in figure 3 appears to have been due to queries of this type, and effectively vanishes with the implementation of these queries.

\subsubsection{Avoiding downtime when indexing tables}

Because DPM needs to write to one of \texttt{dpm\_db} tables for every request it receives, one cannot build indices on the tables in the most straightforward way. MySQL locks InnoDB tables while building indices on them, and this will cause the relevant kind of request on the DPM to fail. We avoided this by the following procedure:
\begin{itemize}
\item Examine the table to determine the highest-numbered row which refers to an ``historical'' request - one which is not undergoing any changes anymore as it is complete and the requestor has finished with it.
\item Clone the historical part of the table to a new table (called, say \texttt{dpm\_req\_copy}), with indices (in these examples, FOO is the last ``historical'' record.
\begin{verbatim}
CREATE TABLE dpm_putfilereq_copy LIKE dpm_put_filereq;
CREATE INDEX status_idx ON dpl_put_filereq_copy;
INSERT INTO dpm_put_filereq_copy SELECT * FROM dpm_put_filereq WHERE rowid < FOO; 
\end{verbatim}
\item In general, the number of ``non-historical'' events in the original table is small. This step may take hours as, if you have a DPM which has been in production for a long time, there are vastly more historical entries than current ones.
\item Once completed, stop the dpm and dpnsdaemon processes on the DPM headnode.
\begin{verbatim}
service dpm stop
service dpnsdaemon stop
\end{verbatim} 
 This is because MySQL cannot rename locked tables, so one has to stop the tables being altered by stopping the services which can write to them.
\item Copy the rest of the rows to the copy table (they will be indexed automatically).
\begin{verbatim}
INSERT INTO dpm_put_filereq_copy SELECT * FROM dpm_put_filereq WHERE rowid > FOO-1; 
\end{verbatim}
\item Switch the table names, so that the ``copy'' is now the ``real'' table.
\begin{verbatim}
RENAME TABLE dpm_put_filereq TO dpm_put_filereq_old
RENAME TABLE dpm_put_filereq_copy TO dpm_put_filereq 
\end{verbatim}
\item Start up dpm and dpnsdaemon processes.
\begin{verbatim}
service dpm start
service dpnsdaemon start 
\end{verbatim}
\end{itemize} 

\subsubsection{Bufferpool size optimisation}

We discovered, after performing these optimisations, that the default MySQL configuration for DPM barely assigns any memory to the InnoDB bufferpool, which is used by the InnoDB engine to cache and buffer common table reads and writes. As a result, the hit-rate on the buffer was around 97\% under some load. (Thus, 3\% of reads hit the disk.) We increased the bufferpool size to half the physical RAM available on the server. The resulting effect on the hit-rate was to increase it to 99.9\%, reducing effective disk load on reads to 1/30th of its previous value. Writes are less easy to optimise in this way, due to the need for fsyncs to be called on write transactions for data integrity. 

\subsection{Post-MySQL optimisation tests}

\begin{figure}
\begin{minipage}{18pc}
\includegraphics[width=18pc]{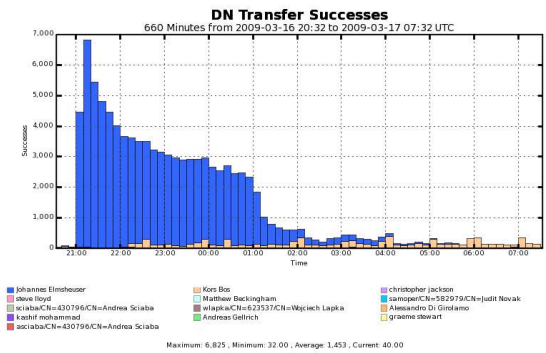}
a) Successful transfers (600 second intervals) by DN
\end{minipage}
\hspace{2pc}
\begin{minipage}{18pc}
\includegraphics[width=18pc]{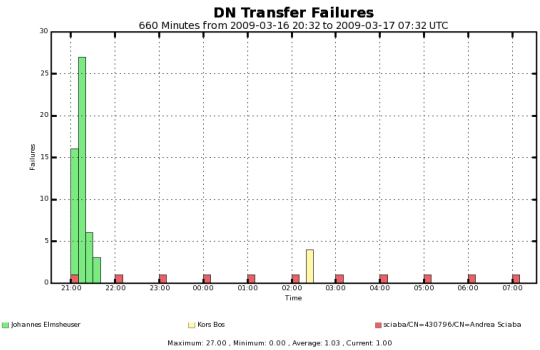}
b) Failed transfers (600 second intervals) by DN
\end{minipage}
\caption{\label{fig5a} DPM statistics for HammerCloud test ``HC193''}
\end{figure}

\begin{figure}
\begin{minipage}{38pc}
\begin{minipage}{18pc}
\includegraphics[width=18pc]{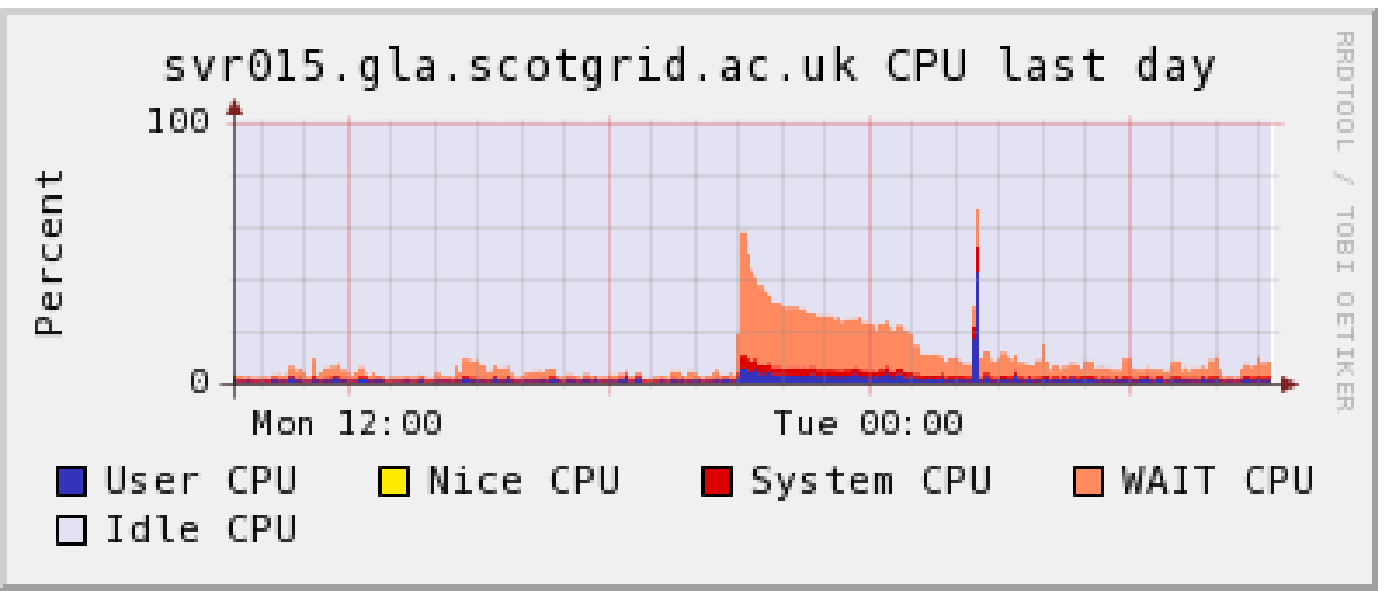}
a) CPU load on MySQL server (percentage of full load)
\end{minipage}
\hspace{2pc}
\begin{minipage}{18pc}
\includegraphics[width=18pc]{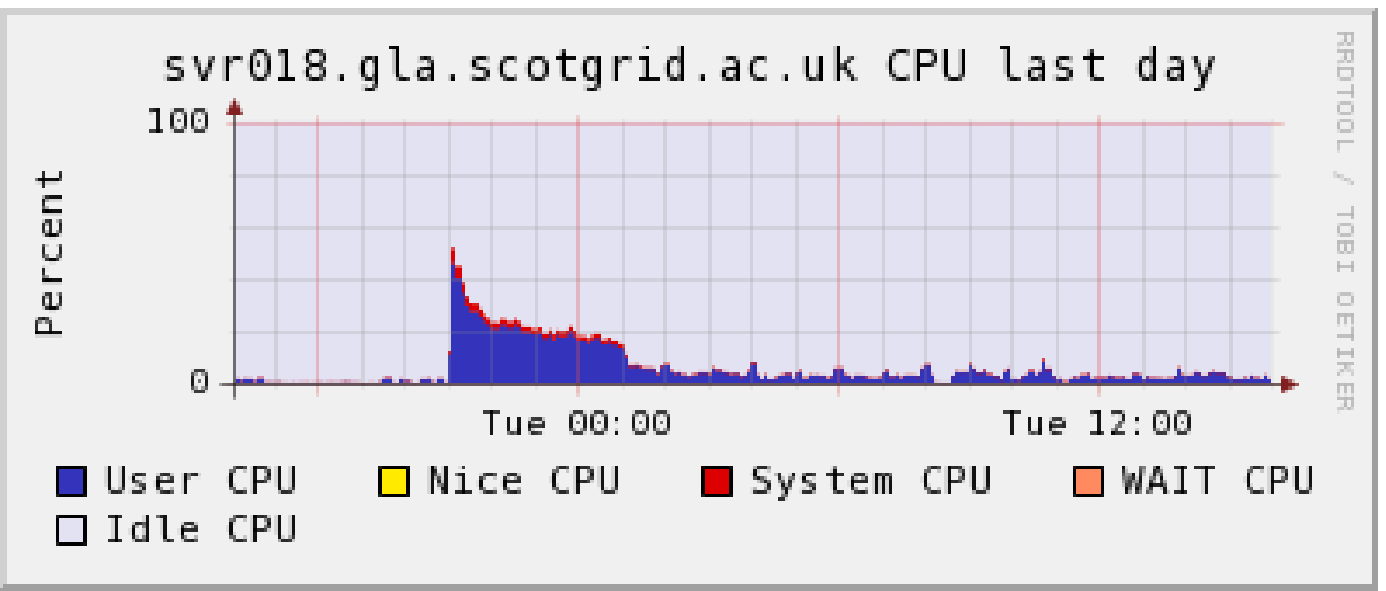}
b) CPU load on DPM services node (percentage of full load)
\end{minipage}
\end{minipage}
\begin{minipage}{38pc}
\begin{minipage}{18pc}
\includegraphics[width=18pc]{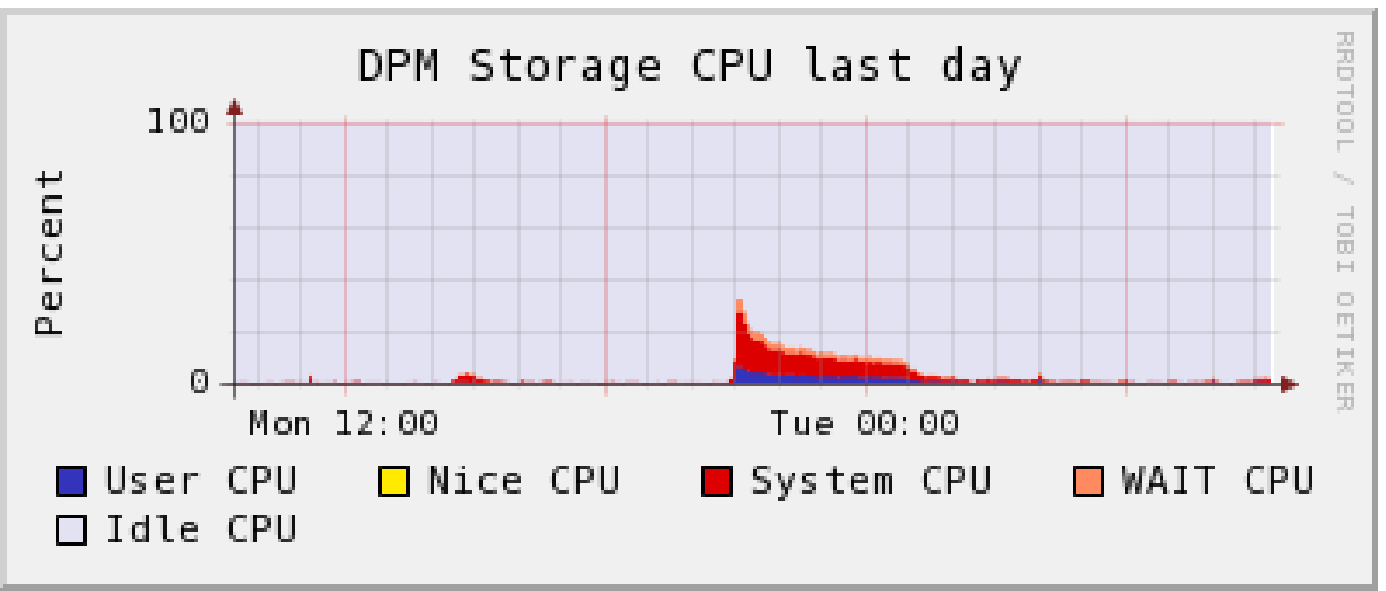}
c) CPU load on DPM pool (average, percentage of full load)
\end{minipage}
\hspace{2pc}
\begin{minipage}{18pc}
\includegraphics[width=18pc]{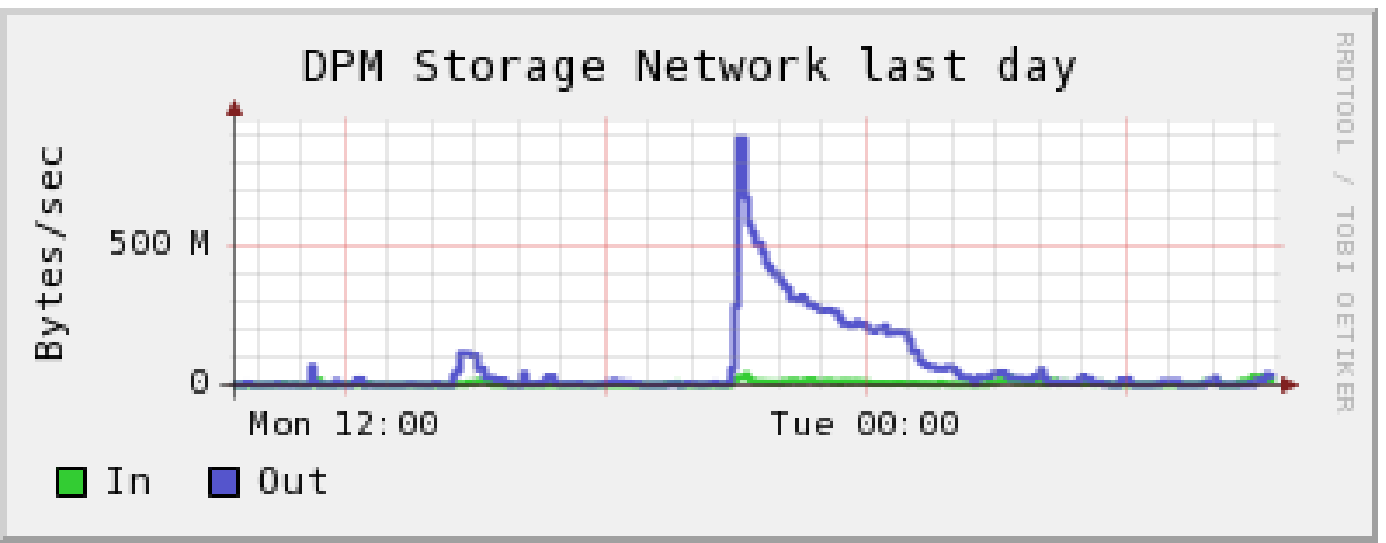}
d) Network load on DPM pool (total, bytes/sec)
\end{minipage}
\end{minipage}
\caption{\label{fig5} Load on services during HammerCloud test ``HC193''.}
\end{figure}

\begin{figure}
\begin{minipage}{10pc}
\includegraphics[width=10pc]{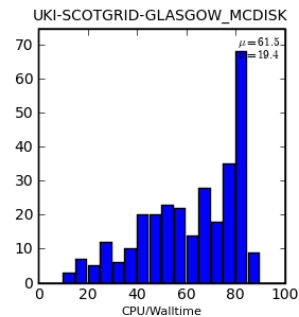}
a) CPU efficiency (cputime/walltime)
\end{minipage}
\hspace{2pc}
\begin{minipage}{10pc}
\includegraphics[width=10pc]{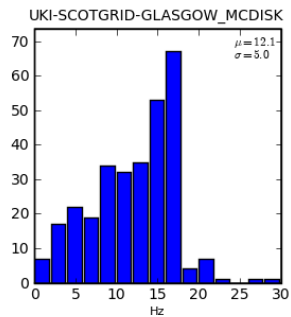}
b) Event rate (events per second per job)
\end{minipage}
\hspace{2pc}
\begin{minipage}{12pc}
\includegraphics[width=12pc]{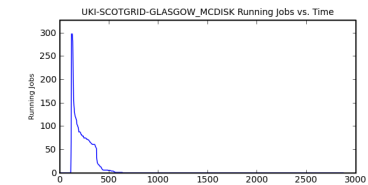}
c) Simultaneous running jobs
\end{minipage}
\caption{\label{fig6} Statistics generated by HammerCloud test ``HC193'' for the Glasgow site.}
\end{figure}

Figures \ref{fig5a} to \ref{fig6} show the performance of the DPM after MySQL optimisation. The load on the MySQL server is clearly somewhat reduced, and significantly less "noisy" than it was previously, and the cpu/io load on the pool nodes still hasn't increased. 
There is a small increase in the successful transfer rate - from 600 to 700 per minute; however, this is not reflected in the event-rate, which appears unexpectedly low (although still better than the unoptimised case). HC135 occurred during a time when the cluster did not have enough free job slots for all analysis jobs to start simultaneously, whilst HC193 arrived on an almost empty cluster. The event-rate appears lower for HC193 because the higher transfer rate is spread over significantly more simultaneous jobs. 
Of concern is the significant increase in transfer failures after optimisation. It appears that this is due to saturation of the network connections of the pool servers.


\section{Conclusions}

By a combination of partitioning of hardware and the application of a small number of tweaks to MySQL configuration, we have significantly improved the performance of the DPM storage infrastructure at UKI-SCOTGRID-GLASGOW with respect to ``typical'' user analysis jobs as represented by HammerCloud tests. 
Further improvements would require upgrading of physical hardware, moving from Gigabit ethernet to 10GigE (or equivalent bandwidth solutions, such as SDR Infiniband), resulting in a significant expense (although it is certain that such upgrades will happen before the LHC is turned on).
User analysis stresses storage in a way that production use does not - requiring transfers of multiple small (AOD) files for each job, with very few pauses between transfers\footnote{ATLAS are now distributing merged AOD files to Tier-2 sites which will reduce the number of \texttt{open()} calls the headnode needs to support and should allow us, with the optimisations in place, to support even more user analysis jobs on the cluster.}. Thus, for a cluster with many fast worker nodes, like UKI-SCOTGRID-GLASGOW, it is important to have as fast a DPM head node as possible. IOwait on the database backend is a significant limiter of total performance, and reducing this significantly affects the total performance. Ultimately, however, you become limited by the capacity of your networking, and the seek rates on the disk servers themselves (as well as the overhead of GSI authentication per request).

\section*{Acknowledgements}
The authors would like to acknowledge the prompt help of Johannes Elmsheuser and Dan van der Ster in scheduling the HammerCloud tests needed to perform this optimisation process.

This work was supported by the GridPP project, funded by the UK Science and Technologies Facilities Council. Stuart Purdie was also supported by the Enabling Grids for E-sciencE, funded by the EU.

\end{document}